# LARP LHC 4.8 GHZ SCHOTTKY SYSTEM INITIAL COMMISSIONING WITH BEAM*

Ralph J. Pasquinelli, Fermilab, Batavia, IL*, U.S.A., Andreas Jansson, ESS, Lund, Sweden,

O. Rhodri Jones, Fritz Caspers, CERN, Geneva, Switzerland

*Abstract*

The LHC Schottky system consists for four independent 4.8 GHz triple down conversion receivers with associated data acquisition systems. Each system is capable of measuring tune, chromaticity, momentum spread in either horizontal or vertical planes; two systems per beam. The hardware commissioning has taken place from spring through fall of 2010. With nominal bunch beam currents of $10^{11}$ protons, the first incoherent Schottky signals were detected and analyzed. This paper will report on these initial commissioning results. A companion paper will report on the data analysis curve fitting and remote control user interface of the system.

## INTRODUCTION

The Schottky system for the LHC was proposed in 2004 under the auspices of the LARP [1] collaboration. Similar systems were commissioned in 2003 in the Fermilab Tevatron and Recycler accelerators as a means of measuring tunes noninvasively. The Schottky detector is based on the stochastic cooling pickups [2] that were developed for the Fermilab Antiproton Source Debuncher cooling upgrade completed in 2002. These slotted line waveguide pickups have the advantage of large aperture coupled with high beam coupling characteristics. For stochastic cooling, wide bandwidths are integral to cooling performance. The bandwidth of slotted waveguide pickups can be tailored by choosing the length of the pickup and slot spacing. The Debuncher project covered the 4-8 GHz band with eight bands of pickups, each with approximately 500 MHz of bandwidth. For use as a Schottky detector, bandwidths of 100-200 MHz are required for gating, resulting in higher transfer impedance than those used for stochastic cooling. Details of hardware functionality are reported previously. [3] [4] [5]

## HARDWARE PERFORMANCE

The hardware was designed with two front-end gain paths 38 dB and 15 dB. This was based on the attempt to measure pilot bunches of $10^9$ protons during initial commissioning. The high gain path is used almost exclusively and has not shown any saturation effects. One hundred dB of instantaneous dynamic range has been achieved. Figure 1 is a measurement at baseband showing this dynamic range for an input signal of +10 dBm to -90 dBm to the signal processor electronics. Pickup motion to reduce common mode signals was not allowed due to possible introduction of an aperture restriction. Future commissioning efforts will include adjustment of a variable attenuator and delay line in one pickup leg to reduce common mode signals. A 0 to 32 dB input step attenuator is utilized to maximize dynamic range under a variety of beam current conditions.

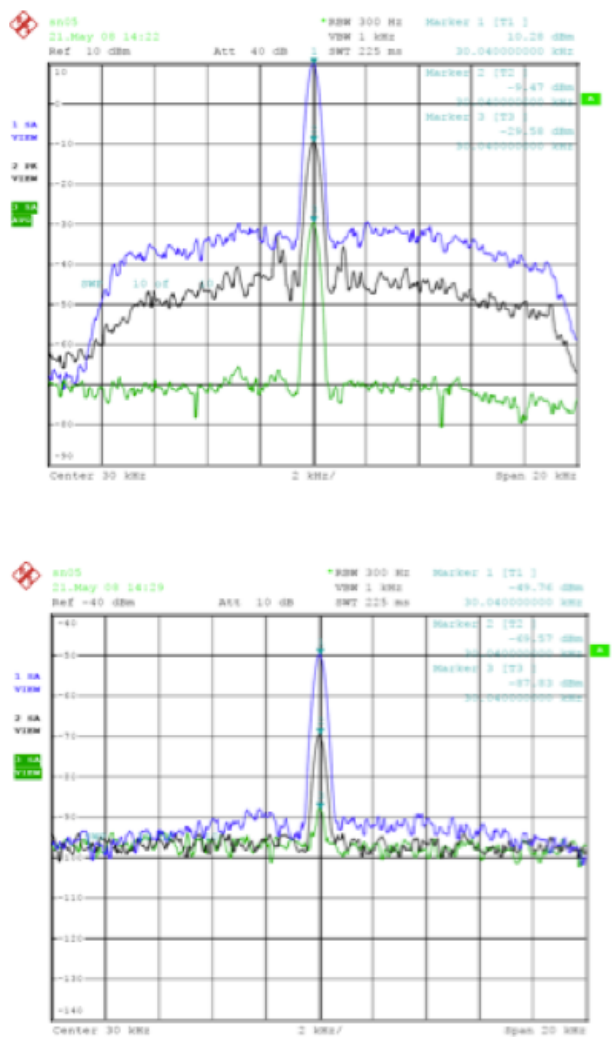

Figure 1: Instantaneous dynamic range of 100 dB in signal processor chassis measured at baseband. Test signal injection of +10 to -90 dBm, 10 dB per division. At the higher signal levels, the 15KHz bandwidth of the crystal filters is evident.

Figure 2 depicts the block diagram of the signal processing chain. Wide bandwidth is maintained until just past the gating switch. From this point, bandwidth is

---

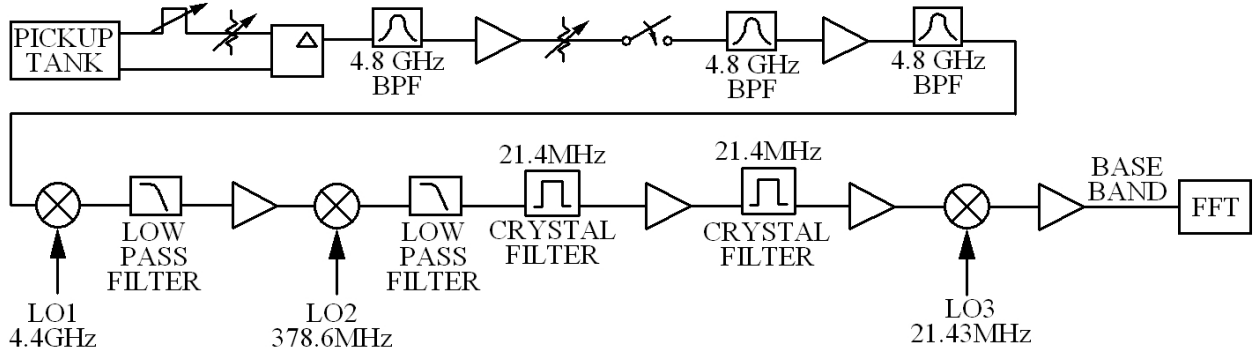

Figure 2: Triple down conversion block diagram of Schottky systems. Common mode rejection adjusted with front-end attenuator and delay line. Fast gate allows for bunch-by-bunch measurements. Base bandwidth 15 KHz.

first reduced to 5 MHz with cavity filters and eventually to 15 KHz with crystal filters. Each revolution band contains all details about tune, chromaticity, and momentum spread. A calibration system allows the injection of a signal to the back termination port of the pickup. This allows for complete gain characterization of the entire processing chain including the pickup. The processor chassis has all three IF frequencies available on the front panel to ease troubleshooting. The 21.4 MHz IF is the only analog signal transmitted via 7/8-inch coax to the surface service building. For emittance measurements utilizing the Schottky, calibration is via comparison to alternate means of emittance measurement.

Development of the processing hardware was carried out at Fermilab. Performance of triple down conversion was a significant improvement over the original single sideband processing electronics utilized in the Tevatron, resulting in a retrofit to this hardware for the Tevatron. Selection of high quality local oscillators with low phase noise is essential as shown in a test done at the Tevatron in Figure 3. The LHC with a revolution frequency of 11 KHz is more sensitive to phase noise for the closely spaced tune lines.

A software team at Fermilab has designed and executed a user friendly control interface allowing gain changes, local oscillator control, and gating settings. The interface also allows for display and continuous monitoring of the Schottky signals, which are then data logged in the CERN control system. [6]

## PROTONS AND LEAD IONS

Tune, chromaticity, and momentum spread measurements for both protons and lead ions have been demonstrated and verified against other instruments measuring these same parameters. Figures 4, 5, and 6 show typical results for protons and lead ions. Note that with protons, the coherent line is sometimes present on the betatron sidebands.

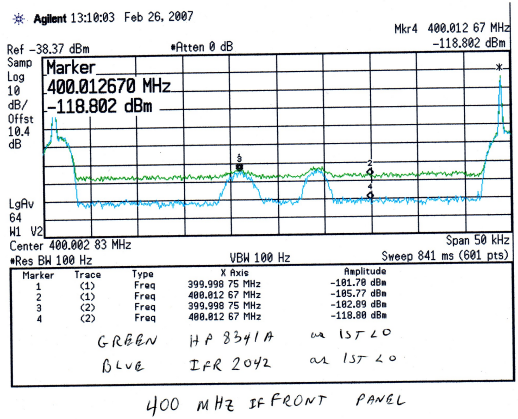

Figure 3. 400 MHz IF signal (front panel) after first down conversion. Sample of Tevatron Schottky signal using two different quality phase noise signal generators as local oscillators.

The resolution capabilities of the CERN FFT system indicate that these lines are synchrotron satellites. These sideband satellites have not been observed in the Tevatron. In contrast with the Tevatron, the LHC Schottky system has also proven to be a valuable measurement tool during the beam energy ramp.

The measurement of chromaticity is particularly useful at injection energy, where the decay of the higher mode dipole component translates into a chromaticity drift, which, if left uncorrected, can lead to negative chromaticity and instability. The Schottky monitor system currently is the only monitor capable of bunch-by-bunch tune measurements and was used extensively during studies of the electron cloud phenomena observed when testing 50ns bunch spacing operation towards the end of the 2010 run.

Lead ion signals were significantly cleaner than proton signals and seemed to suffer less from coherent contributions to the transverse spectra, allowing an on-line measurement of tune and chromaticity throughout the accelerator cycle. The proton spectra suffered from the added complication of the RF longitudinal blow-up. The beam blow up being performed purposely to maintain a relatively long bunch length throughout the ramp. This produced a wide longitudinal component at the revolution line, which completely swamped the transverse Schottky signals. As has been observed at the Tevatron, the cleanest spectra were obtained several tens of minutes after initiating collisions.

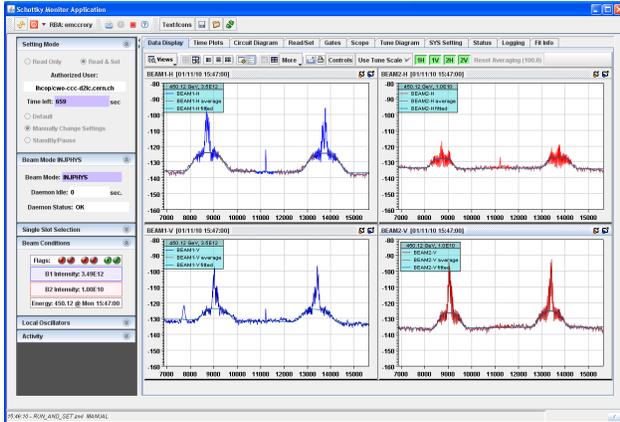

Figure 4. LHC Schottky transverse sideband signals for $3.5 \times 10^{12}$ protons at 450 GeV injection energy. Coherent lines observed with synchrotron satellites are excluded from tune fit. Frequency scale is zoomed in not showing the revolution line.

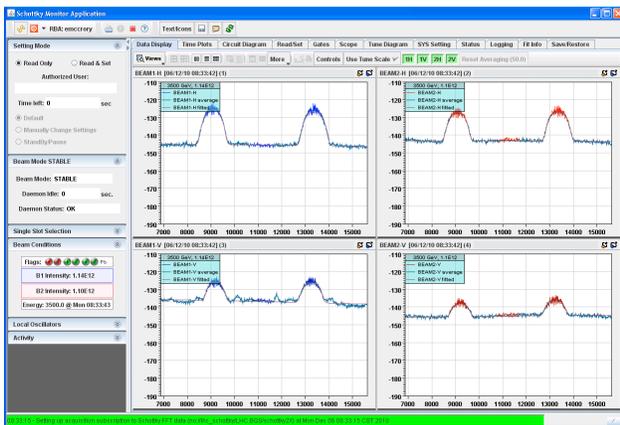

Figure 5. LHC Schottky transverse sideband signals for $1.1 \times 10^{12}$ lead ions at 3.5 TeV.

An added advantage of the LHC Schottky monitor, with its high detection frequency at 4.8 GHz, is its immunity to the effects of the transverse damper. The transverse damper system is being used at a relatively high gain to combat instabilities and emittance blow-up. The damper induced interference to the standard tune system is not observed with the Schottky system. The LHC Schottky system is therefore being considered as a serious alternative to the standard tune system for providing data to auto tune feedback. Implementation for tune feedback will require a faster tune acquisition than is currently possible with existing data acquisition hardware plus a robust measurement throughout the ramp, even in the presence of longitudinal blow-up.

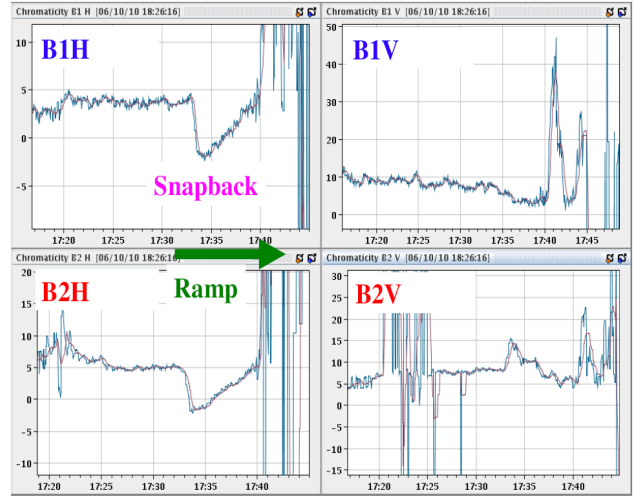

Figure 6. Chromaticity vs. time for protons. Control panel from LHC Schottky application program.

## CONCLUSIONS

Commissioning efforts continue with the start of the 2011 run. Automated bunch-by-bunch gating measurements, data logging of parameters, and inclusion in daily operation of the LHC are planned.

## REFERENCES

[1] http://www.uslarp.org/

[2] D. McGinnis, "Slotted Waveguide Slow-Wave Stochastic Cooling Arrays", PAC '99, New York

[3] R. J. Pasquinelli et al, "A 1.7 GHz Waveguide Schottky Detector System", PAC 2003, Portland, Oregon.

[4] A. Jansson et al. "Experience with the 1.7 GHz Schottky Pickups in the Tevatron", EPAC 2004, Lucerne, Switzerland.

[5] F. Caspers et al. "The 4.8 GHz LHC Schottky Pickup System", PAC 2007 Albuquerque, New Mexico.

[6] J. Cai et al, "High Level Software for 4.8 GHz LHC Schottky System", PAC 2011 MOP218, New York, NewYork